\documentstyle[12pt]{article}
\author{Serge Galam and Alain Mauger\\
Acoustique et Optique de la Mati\`{e}re Condens\'{e}e\footnotemark[1]\\
Tour 13 - Case 86, 4 place Jussieu, 75252 Paris Cedex 05, France\\[1ex].}
\title{Reply to the comment``Universal Formulae for Percolation Thresholds"  }
\addtocounter{footnote}{0}
\footnotetext{Laboratoire associ\'{e} au CNRS (URA n$^{\circ}$ 800) et
\`{a} l'Universit\'{e}
P. et M. Curie - Paris 6}
\addtocounter{footnote}{1}
\date{Phys. Rev. {\bf E}\underline {55}, 1230 (1997)}

\begin{document}
\maketitle
\baselineskip 3.3ex
\footskip 5ex
\parindent 2.5em
\abovedisplayskip 5ex
\belowdisplayskip 5ex
\abovedisplayshortskip 3ex
\belowdisplayshortskip 5ex
\textfloatsep 7ex
\intextsep 7ex
\begin{center}
{\em PA Classification Numbers:\/} 64.60 A, 64.60 C, 64.70 P\\
\end{center}

%%%%%%%%%%%%%%%%%%%%%%%%%%%%%%%%%%%%%%%%%%%%%%%%%%%%%%%%%%%%%%%%%%%%%%%
\begin{abstract}

In a recent paper, we have reported a universal power law for both site
and bond percolation thresholds for
any lattice of cubic symmetry. 
Extension to anisotropic lattices is discussed.
\end{abstract}
%%%%%%%%%%%%%%%%%%%%%%%%%%%%%%%%%%%%%%%%%%%%%%
\newpage

In Ref. 1, we have found a power law for both site and bond percolation
thresholds, which writes
\begin{equation}
p_c=p_0[(d-1)(q-1)]^{-a}d^{\ b}
\end{equation}
with $d$ the space
dimension and $q$ the coordination number.
For site dilution $b=0$ while
$b=a$ for bond dilution. Then a class of lattices is defined by the set
of parameters $\{p_0; \ a\}$.
One class includes two-dimensional triangle, square
and honeycomb lattices, characterized by
$\{p_0=0.8889; \ a=0.3601\}$ for
site dilution and by
$\{p_0=0.6558; \ a=0.6897\}$ for bond dilution. Two-dimensional Kagom\'{e}
and all
 other lattices of cubic symetry (for
$d\geq 3$) constitute the second class,
 characterized by
$\{p_0=1.2868; \ a=0.6160\}$ and
$\{p_0=0.7541; \ a=0.9346\}$ for sites and bonds respectively.
At high dimensions a third class for hypercubes (sc and fcc) is found,
which recovers the infinite Cayley tree limit.

In above comment van der Marck reports the interesting observation 
that the stacked triangular lattice
(also called hexagonal lattice) with the lattice parameters $a=b=c$ ($d=3$, $q=8$) does not
fit into the second class. In particular, the percolation thresholds reported
are different from those associated to the $d=3$, $q=8$ bcc lattice.

This is indeed an interesting observation which,
however, does not contradict our previous work for the following reason.
Within a given class, the percolation threshold after Eq. (1) depends only
on $d$ and $q$, which implicitely requires that the q nearest neighbors of any
site are equivalent. This is indeed the case in all lattices we have mentioned
in the definition of the classes, but that is not the case of the
stacked triangular lattice which is anisotropic. There a lattice site has 6 equivalent
nearest neighbors in the $a,b$ plane (bonding angle is $60^0$) and 2
non-equivalent sites along the $c$ axis (bonding angle is $90^0$).

Actually,
the percolation threshold of an anisotropic lattice must depend on the degree
of anisotropy. This can be viewed on the stacked triangular lattice, if we note that
this lattice is defined by lattice parameters $ a=b\ne{c}$ (the case $a=c$
considered in the comment is only a very particular case).
 Then in the limit where the $c$ parameter goes to infinity, one is left
with $ab$ planes which will become decoupled for physical systems with finite
ranges of interaction. Therefore, the percolation
threshold of the stacked triangular lattice must depend on the ratio $c/a$ and the
percolation threshold will shift continuously from the numerical values
given by van der Mark in the particular case $(c/a = 1, q=8)$, to those of the
triangular lattice in $d=2$, in the limit $c/a \rightarrow \infty$. Note in this
limit one recovers an isotropic lattice with $q=6$ instead of $q=8$. 

However it should be stressed that above interpolation should not be taken literraly. 
Percolation thresholds do depend on site connectivity and not on length between them. 
Then, if one wishes to
generalize Galam-Mauger formula, anisotropy should be taken into
account by replacing the $q$ parameter by some effective value between
 $q=8$ and $q=6$. Indeed, we found that a unique value of $q=6.65$ reproduces
within the second universality class at $d=3$
both percolation thresholds 0.2614 (site) and 0.1875
(bond) in agreement with the result of Van der Marck for the
stacked triangular lattice with $c=a$ which are  0.2623 and 0.1859
respectively.

According to these consideration, we can now make a discussion of the limits
of validity for the Galam-Mauger formula, which was lacking in Ref. 1.
Along the stacked triangular lattice case, one may also construct
 anisotropic percolation problems by
 having, for instance, two different bond probabilities in the two different
lattice directions
of the square lattice. Directed percolation would be another example. All these
problems are more complicated that the isotropic percolation problem considered
in Ref. 1, and were not considered in this prior work.

Extension of
the formula, in view of above discussion, seems however possible, if one replaces
 $q$ by an effective value. However this value has
to be determined for each case, depending on the nature and strength of the
anisotropy. Nevertheless Galam-Mauger
formula preserves a capacity of prediction. Knowledge of one (either
site or bond)
percolation threshold allows the determination of the effective value of
$q$ for each
anisotropic percolation problem investigated. Then this value
can be used to estimate the other percolation threshold. Otherwise the direct
estimate of the effective $q$ is required to yield both percolation thresholds.

Note that Galam-Mauger formula actually applies not only to 
lattices with cubic
symetry
investigated in Ref. 1, but also to all isotropic lattices in general. This
can be illustrated with the  hexagonal compact (hcp) lattice. This is
actually not a Bravais
lattice, because, on a topologic view point, it is a simple hexagonal lattice
with two atoms per unit cell. However each atom in this structure has $q=12$
nearest neighbors with the same bonding angle for each of them. We are then
in the isotropic case with $d=3$, $q=12$ so that we
predict the same percolation thresholds as in the case of fcc lattice
at d=3, namely 0.192 and 0.117 for site and bond percolation thresholds
respectively according to Galam-Mauger formula.

Percolation thresholds
for the hcp were given by V. K. S. Shante and S. Kirpatrick [2]
two decades ago, giving 0.204 (site) and 0.124 (bond). These authors
report for the fcc site and bond percolation thresholds 0.199 and 0.125
respectively. Comparison with modern values available for the fcc shows that
their last digit is not accurate. We can then conclude that the agreement
between fcc and hcp percolation thresholds is good and actually much better
than hexagonal and bcc ones as expected.

%%%%%%%%%%%%%%%%%%%%%%%%%%%

\subsection*{Acknowledgments.}
We would like to thank Dietrich Stauffer for
very stimulating discussions.

%%%%%%%%%%%%%%%%%%%%%%%%%%%%%%%%%%%%%
\vspace{3.0cm}
{\LARGE References}\\ \\
1. {\sf S. Galam and A. Mauger}, Phys.Rev.E
\underline {53},
2177 (1996)  \\
2. {\sf V.K.S. Shante and S. Kirpatrick}, Adv. Phys. \underline {20}, 326
(1971) \\

%%%%%%%%%%%%%%%%%%%%%%%%%%%%%%%%%%%%

%%%%%%%%%%%%%%%%%%%%%%%%%%%%%%%%%%%%%
\end{document}